\documentclass[conference]{IEEEtran}
\IEEEoverridecommandlockouts

\usepackage[utf8]{inputenc} 
\usepackage[english]{babel}
\usepackage{amsmath,amssymb,amsfonts}
\usepackage{graphicx}
\usepackage{svg} 
\usepackage{textcomp}
\usepackage{xcolor}
\usepackage{pgfplots}
\usepackage{pgfplotstable}
\pgfplotsset{compat=1.18} 

\usepackage[hidelinks]{hyperref}

\usepackage{tcolorbox}
\usepackage{url}
\usepackage{tikz}
\usepackage{float}
\usepackage{stmaryrd}
\usepackage{setspace}
\usepackage{caption} 
\usepackage{subcaption}
\usepackage{stfloats}
\usepackage{comment}
\usepackage{soul}
\usepackage{csquotes}
\usepackage{algorithm}
\usepackage{algpseudocode}

\usepackage{cite} 

\usepackage{booktabs} 

\usetikzlibrary{patterns}
\usetikzlibrary{positioning}

\definecolor{babyblueeyes}{rgb}{0.63, 0.79, 0.95}
\definecolor{bluecurve}{RGB}{77, 88, 208}
\definecolor{chad-green}{RGB}{79, 159, 98}

\NewDocumentCommand\refFig{mm}{Fig.~\ref{fig:#1}#2}

\def\BibTeX{{\rm B\kern-.05em{\sc i\kern-.025em b}\kern-.08em T\kern-.1667em\lower.7ex\hbox{E}\kern-.125emX}}

\begin{document}

\title{Training LLMs for Generating IEC 61131-3 Structured Text with Online Feedback}

\author{
    \IEEEauthorblockN{
        Aaron Haag\IEEEauthorrefmark{1}, 
        Bertram Fuchs\IEEEauthorrefmark{1}, 
        Altay Kacan\IEEEauthorrefmark{1}, 
        Oliver Lohse\IEEEauthorrefmark{1}
    }
    \IEEEauthorblockA{
        \IEEEauthorrefmark{1}Siemens Technology Department, Munich, Germany\\
        Email: \{aaron.haag, bertram.fuchs, altay.kacan, oliver.lohse\}@siemens.com
    }
}


\maketitle

\makeatletter
\def\ps@IEEEtitlepagestyle{
  \def\@oddfoot{\mycopyrightnotice}
  \def\@evenfoot{}
}
\def\mycopyrightnotice{
  {\footnotesize
  \begin{minipage}{\textwidth}
  \centering
 This work has been submitted to the IEEE for possible publication. Copyright may be transferred without notice, after which this version may no longer be accessible.
  \end{minipage}
  }
}

\begin{abstract}
IEC 61131-3 Structured Text (ST) is a widely used programming language for programmable logic controllers (PLCs) in automation systems. However, generating ST code with LLMs poses unique challenges due to limited data in public training datasets and the complexity of ST language syntax. This paper proposes an approach to fine-tune LLMs for the generation of ST code that leverages a preference-based learning method through an online process involving compiler feedback and evaluation from an LLM-based ST expert. In this framework, the model is iteratively refined and generates new training samples, which are subsequently evaluated by a compiler for syntactical correctness and by a specialized LLM that excels at assessing semantic accuracy, though it is not optimized for code generation itself. This approach results in marked improvements for the trained LLM, leading to higher compilation success rates and better semantic precision. As a result, the framework proves highly suitable for industrial automation applications and outperforms state-of-the-art models.
\end{abstract}




\section{Introduction}


We are entering a new era of code intelligence, where AI-driven copilots are transforming how developers interact with code. These tools, powered by Large Language Models (LLM), streamline complex programming tasks and significantly enhance productivity \cite{barke2022groundedcopilotprogrammersinteract}. However, as we approach a potential depletion of high-quality internet data by 2026, the necessity of synthetic data for model training will increase \cite{liu2024bestpracticeslessonslearned}. This shift presents significant challenges, particularly for niche programming languages such as IEC 61131-3 Structured Text (ST), which is prevalent in Programmable Logic
Controller (PLC) engineering and industrial automation systems \cite{siemensdocst}. To ensure that LLMs provide value in such specialized domains, it is essential that these models are aligned with expert expectations and values, optimizing their output for real-world needs \cite{bubeck2023sparksartificialgeneralintelligence,  ouyang2022traininglanguagemodelsfollow}. 


Reinforcement Learning from Human Feedback (RLHF) \cite{christiano2023deepreinforcementlearninghuman}, a pioneering LLM alignment method, trains a reward model (RM) using pairwise human preferences and then optimizes the policy through reinforcement learning (RL) based on the RM. Recently, Direct Alignment from Preferences methods (DAP) have emerged as popular alternatives to RLHF \cite{rafailov2024directpreferenceoptimizationlanguage}, \cite{ethayarajh2024ktomodelalignmentprospect}. Unlike RLHF, DAP methods directly update the language model using pairwise preference data, simplifying the alignment process because there is no need to train a dedicated reward model. In the original DAP approach \cite{rafailov2023directpreferenceoptimizationlanguage}, preference datasets are gathered prior to training and are based on expert feedback. With LLMs growing in size and complexity, this method can become prohibitively expensive if human experts label IEC 61131-3 conform generated code. Therefore, innovative alignment methods beyond direct human feedback are needed to fully leverage these models.

 
In response to this challenge, this work utilizes synthetic experts in an iterative online approach based on syntactic compiler feedback and a dedicated semantic LLM expert. To overcome the data scarcity of PLC code samples, the proposed framework exclusively exploits preexisting user prompt and code pairs derived from the APPS dataset \cite{hendrycks2021measuringcodingchallengecompetence} for SFT, as well as additional prompts that are subsequently used in an online finetuning scheme similar to Online AI Feedback (OAIF) \cite{guo2024direct}. This eliminates the need for large code datasets. The approach iteratively refines the model and incorporates feedback on both syntactic and semantic correctness, ensuring progressively better outcomes.



We base our proposed method on three concepts to overcome data scarcity in the ST code generation domain. First, there exists a performance margin between code generation and code quality judgement. Datasets used for training State-of-the-Art GPTs include only a small number of ST code samples, or none at all. Even though these models may not be able to generate ST code themselves, they are able to judge ST code samples due to semantic similarities with other languages like Pascal, which they were trained on extensively. Similarly to concepts like OAIF \cite{guo2024direct} or \cite{chen2024selfplayfinetuningconvertsweak} we use the margin between generation and evaluation performance of SOTA LLMs to enhance ST code generation. While DPO typically applies to static offline preference use, recent studies found that DPO also benefits from iterative online training \cite{xiong2024iterative, yuan2024self, xu2024dpo}.


Second, using compiler feedback allows an accurate mapping from a diverse distribution of code samples to a binary compile label for each intent input \cite{wang2022compilable}. This potentially enables training on an infinite amount of synthetically generated and syntactically flawless code samples beyond the syntax judgment performance of LLM judges.


Third, leaning on the concept of natural selection, a set of many code samples is generated instead of only code sample pairs. This enables the syntactic (compiler feedback) and semantic judge (LLM) to explore a broader distribution of samples \cite{brown2024large}. It was shown that a high number of samples results in improved output quality \cite{hassid2024larger, brown2024large}. Hence, we believe that a sufficient number of code samples has to be generated to find a syntactically and semantically flawless code sample.


The remainder of this paper is organized as follows: Section II reviews state-of-the-art methods. Section III introduces the proposed method. Section IV presents simulation results and discusses the future potential of the approach. Finally, Section V concludes the paper.

\begin{figure}[!h]
    \begin{center}
        \includegraphics[width=\columnwidth,keepaspectratio]{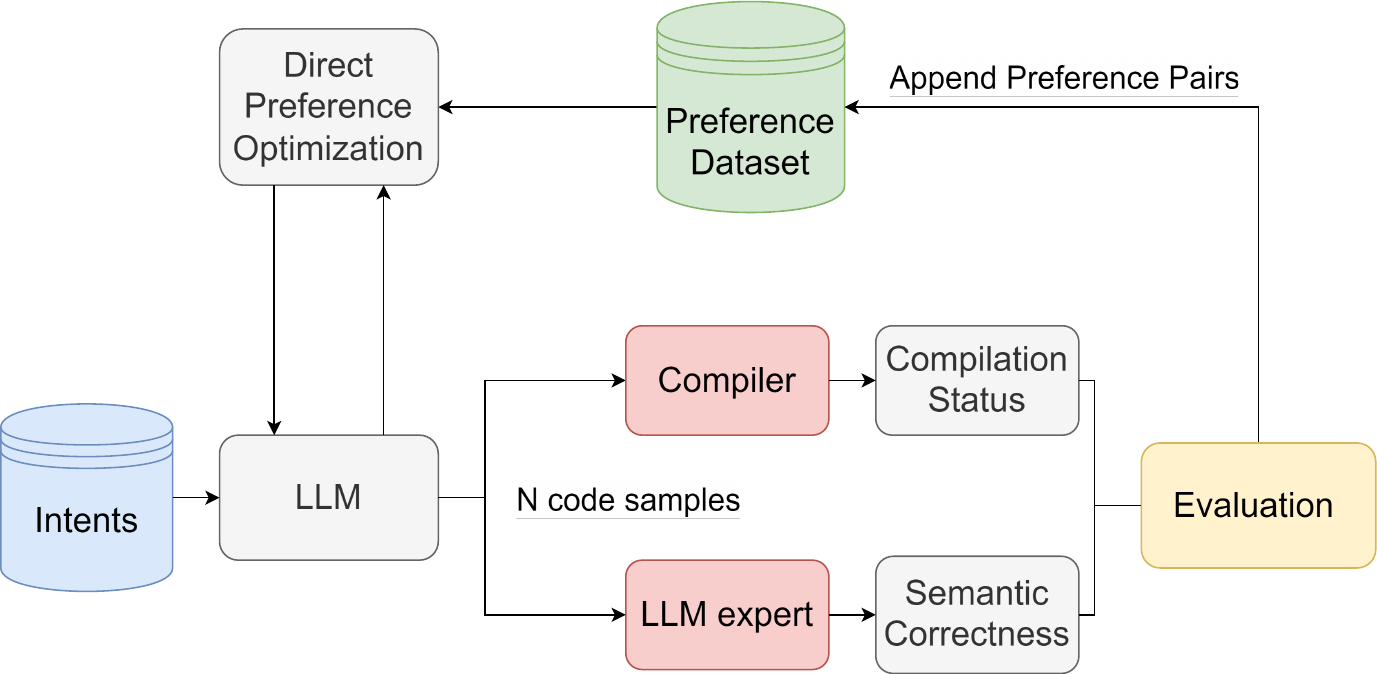}
        \caption{Overview of the DPO fine-tuning architecture, using a combination of feedback from a compiler and LLM experts.}
    \label{fig:pipeline}
    \end{center}
\end{figure}


\section{State-of-the-Art Methods}

Before giving details on the proposed methodology, we investigate state-of-the-art methods for LLMs for code generation, OAIF and compiler feedback.


\subsection{LLM-based Code Generation}

Initial approaches for LLM-based code generation took problem statement prompts as input and directly generated code outputs conditioned on the prompts \cite{chen2021evaluating}. More recent techniques, such as chain-of-thought prompting \cite{wei2023chainofthoughtpromptingelicitsreasoning} focus on breaking down tasks into modular or pseudo-code steps to improve planning and minimize errors. Furthermore, retrieval-based strategies incorporate similar problem samples and code solutions to guide the code generation process \cite{parvez2021retrievalaugmentedcodegeneration}. While these objectives are effective for natural language tasks, they fall short in fully capturing the complex syntactic and semantic rules of programming languages, often resulting in issues like syntax errors and type mismatches. Additionally, the success of these models in code generation is hindered by the need for large domain-specific datasets, which are scarce for specialized fields like industrial PLCs. As a result, LLM-generated code often fails to pass test cases and lacks built-in mechanisms for automatic detection and correction of syntactic and semantic errors \cite{ridnik2024codegenerationalphacodiumprompt}.

\subsection{Recent advances in LLM Alignment}

Generative Pre-trained Transformer (GPT) models show great potential to generate PLC code based on domain-specific commands \cite{10275411}. However, a significant challenge in this area remains the limited availability of data, which complicates effective model training and hinders performance. To address these challenges, the LLM4PLC project fine-tuned an open source model using publicly available PLC libraries such as OSCAT, demonstrating significant improvements in PLC code generation \cite{Fakih_2024}. Despite recent advances, models still struggle with first-attempt code accuracy, typically requiring multiple iterations to resolve compilation errors. For effective use of code in industrial applications, further refinements are needed.

An approach that greatly improves code quality is Reinforcement Learning for Human Feedback (RLHF). In RLHF, human evaluators provide preference data by comparing pairs of model outputs. This data is used to train a reward model (RM), which is incorporated into the LLM model training process \cite{ouyang2022traininglanguagemodelsfollow}. Recently, DAP methods have gained popularity as alternatives to RLHF. DAP methods, such as Direct Preference Optimization (DPO) \cite{rafailov2023directpreferenceoptimizationlanguage}, update the target policy directly from preference pairs by training models with a preference model that simulates the evaluation of human feedback. In real-time, labelers compare model outputs to ensure consistency, but gathering large-scale annotations from human labelers is both expensive and logistically challenging. In most real-world applications, due to the high cost and complexity of collecting pairwise preferences, the preference dataset is typically compiled before aligning a language model and remains fixed throughout training. Gathering real-time preferences on new responses is typically impractical due to the absence of a humans-in-the-loop for continuous feedback. As a result, the fixed dataset makes all preference data static, preventing the policy from receiving feedback on its own generated output during the alignment process. Relevant works, such as Coarse-Tuning \cite{jain2023coarsetuningmodelscodereinforcement}, CodeRL \cite{le2022coderlmasteringcodegeneration}, StepCoder \cite{dou2024stepcoderimprovecodegeneration}, and MapCoder \cite{islam2024mapcodermultiagentcodegeneration} utilize compiler feedback to improve model training. Recent advances use DPO teratively, e.g. \cite{rafailov2023directpreferenceoptimizationlanguage}\cite{xiong2024iterative}.


\subsection{LLM-based Structured Text Code Generation}

Structured Text is one of the programming languages defined in the standard IEC 61131-3 for programmable logic controllers and has been designed for automation and control tasks. As the necessity of LLMs able to generate ST code is prevalent, several groups have developed methods for ST code generation, recently. Multiple base LLM models like GPT-3 and fine-tuned GPT-J have already been finetuned for generating ladder logic, function block diagrams, and ST code \cite{tran2024generating}. The models were found to generate basic PLC code, but struggled with more complex logic and integration with real-world PLC hardware and libraries. Another group showed that retrieval-augmented ST code generation integrates function blocks from libraries like OSCAT into the LLM-generated code \cite{koziolek2024llm}. LLM4PLC further introduced syntax checkers and model verifiers and utilized fine-tuning with Low-Rank Adaptations (LoRAs) to adapt LLMs for ST code generation. The applied methods showed a significant improvement in ST code generation performance, evaluated by domain experts \cite{fakih2024llm4plc}. The summarized papers were all published in 2024 and demonstrate the growing industrial interest for ST code generation. Despite these efforts, improvements are required to meet industry requirements and shrink the margin between finetuned models for ST code and SOTA models for frequently used programming languages like Python.



\section{Method}
This section describes the two subsequent stages for fine-tuning the LLM with special emphasis on the online-dpo method.

\subsection{Supervised Fine-Tuning}

Supervised Fine-Tuning (SFT) adapts a pre-trained language model (LLM) $\theta_0$ for domain-specific tasks. During SFT, the model is trained on a set of instruction-response pairs $D_\text{sft} = \{(d_1, d^{r}_1), \dots, (d_m, d^{r}_m)\}$. This step ensures the model can follow instructions accurately and apply specific syntax, such as ST. SFT bridges the gap between general language understanding from pre-training and the specialized requirements of downstream tasks by maximizing the likelihood of generating the correct response $d^{r}$ (i.e. ST control code) given the instruction $d$ (i.e. code intent), as described in Equation~\ref{eq:sft}.

\begin{equation}
    \mathcal{L}_\text{SFT} = \arg\max_{\theta_0} \sum_{(x, y) \in D_\text{sft}} \log \pi(y|x)
    \label{eq:sft}
\end{equation}

SFT is conducted as the first step of fine-tuning, using an instruction-response dataset consisting of user intents and corresponding code. 
This assumes that a sufficient number of pairs is available. For ST, there are only a few public datasets (e.g., OSCAT \cite{OSCAT2023}). 
The synthetic generation of such datasets can be time-consuming and expensive, especially if the model has limited understanding of code.

\subsection{Direct-Preference-Optimization with AI and Compiler-Feedback}

DPO (Direct Preference Optimization) is a widely used technique for model alignment based on expert feedback. In contrast to SFT (Supervised Fine-Tuning), it utilizes an instruction and preference triple rather than an instruction-response pair. This means that instead of requiring a code example, it only needs to be determined which output from a model, based on a given prompt, is preferred over another. This approach has the advantage of exploiting formal evaluation methods, such as unit tests for code as an expert. An actuall reference code sample is not needed.
Classical offline DPO uses static preference datasets and conducts a single training round on them [19]. These datasets can be generated in various ways, such as by querying a better model (a form of knowledge extraction) or by using the model being trained. If the model lacks sufficient performance, sampling methods (e.g., [smaller models source]) or self-correction approaches based on feedback could be employed [source]. However, this has the drawback of requiring a large number of samples to obtain high-quality positive preference pairs, or it may lead to a distribution shift between the preference samples and the actual model, which could result in degraded learning performance [source?].
To achieve continuous improvement of the model, we employ an iterative online variant of DPO [29][30]. This newer DPO training technique ensures that the preference pairs remain aligned with the model’s distribution, also allowing the model to benefit from its own improvements in generating positive samples [source]. In this approach, the model being aligned generates samples, which are then used as preference pairs for fine-tuning through DPO. This process is repeated iteratively: the updated model generates new outputs, which are again evaluated by the expert to create new preference samples for further fine-tuning. This has the advantage that only instructions and an expert for the preference collection are needed for training, rather than instruction-response pairs with code, which are often scarce in the case of SFT for code.

\subsection{Description of preference collection}
The formal description of the procedure is as follows.

Let the instruction set be defined as:
$\mathcal{X} \triangleq \{x_1, x_2, \dots, x_M\}
$
where \(M = |\mathcal{X}|\) denotes the cardinality of the set. 

For each iteration $i \in \{1, \dots, I\}$, a subset $X_i \triangleq  \{X_{i,1},\dots,X_{i,N}\}$ is drawn from $\mathcal{X}$: $ X_i \sim \pi_i(\mathcal{X}),
$
where $\pi_i(\mathbf{X})$ represents the selection strategy applied during iteration $i$ and $N$ is the number of drawn instructions.
For each instruction, a number of $T$ code responses from the model $\theta_i$ is sampled, the set is denoted by $Y_i \triangleq \{Y_{i,1},\dots,Y_{i,T}\}$.
In this paper, two experts are queried to build preferences over the responses. The first expert is the compiler, which checks the syntactic correctness. The second expert is an LLM, which is used as a soft substitute for unit tests. 
The experts are combined to generate the preference pairs in the following way. Each generated code sample in $Y_i$ is tested on the compilation. 
The samples that compile, are checked by the semantic expert using a LLM. The samples that do not compile, are treated as negative samples. The LLM as semantic expert then labels the compiling samples again as positive or negative, depending on their semantics. All positive samples are then combined with all negative samples. In case no samples are positive after the compile check, all samples are given to the LLM and labeled. 

The combination of the respective expert feedback ultimately results in a preference dataset $D_i$, where $Y_w \succ Y_l$:

\[
D_i = \{(X_{i,n}, y_{w,n}, y_{l,n})\} \forall n\in\{1,\dots,N\},
\].

\subsection{Direct-Preference-Optimization Fine-Tuning}

After collecting the prefence dataset $D_i$ in the $i$th iteration, a model training leverging DPO is conducted and the model parameters are updated to $\theta_{i+1}$.
The overall process is depicted in Fig. \ref{alg:llm_training}.

\begin{algorithm}[!h]
\caption{Fine-Tuning algorithm to train a LLM for ST code generation}
\label{alg:llm_training}
\textbf{Input:}
\begin{itemize}
    \item Number of training iterations $I$
    \item Instruction dataset $\mathcal{X}$
    \item Supervised dataset $\mathcal{D_{\text{sft}}}$
    \item Code generation model $\theta_{0}$ 
    \item Structured Text experts (semantic and syntax experts)
\end{itemize}
\begin{algorithmic}[1] 
\State Initialize $\theta$ with supervised fine-tuning (SFT) to get $\theta_0$
\For{$i \gets 1$ to $I$}
    \State Sample $N$ ST intents $X_i \sim \pi_i(\mathcal{X})$
    \State Generate $T$ ST code responses from $\theta_{i-1}$ for each intent
    \State Leverage expert feedback to construct preference dataset $D_i$
    \State Update model $\theta_i$ using $D_i$ via DPO
\EndFor
\end{algorithmic}
\textbf{Output:} Aligned code generation model $\theta_I$
\end{algorithm}

\section{Simulation Results and Discussion}
This section presents the simulation results that illustrate the effectiveness of the proposed method.
\subsection{Performance Metrics}
To quantify the performance of the method, we define evaluation metrics appropriate for the ST experts which are abbreviated from \cite{fakih2024llm4plc}\cite{chen2021evaluating}.
\begin{itemize}
    \item[$\bullet$] \textbf{Compilation rate, $P_{c}$} –  The empirical probability that a code segment \(C_i\) successfully compiles. This metric reflects the likelihood that the code passes syntax and type-checking stages without errors.
    \item[$\bullet$] \textbf{Semantic correctness rate, $P_{s}$} – The empirical probability that \(C_i\) is semantically accurate. The primary criterion is the quality of the implementation and its alignment with the intent.
    \item[$\bullet$] \textbf{Joint semantic correctness and compilation rate, $P_{j}$} – The empirical joint probability that a generated code sample is both syntactically valid and semantically correct. This metric reflects the overall success in passing both compilation and semantic validation processes.
\end{itemize}

\noindent These customized metrics provide evaluation for the quantitative aspects of the method.

\begin{figure}
    \begin{tcolorbox}
        \textbf{SYSTEM}: ``You are an expert in PLC Structured Text (ST) programming as per IEC 61131-3 standards. Your task is to evaluate the given code based on its logical, semantic, and syntactic correctness. You can tolerate minor mistakes or errors.

IMPORTANT: Use the following format for your responses: ``[0] [1] [1]", where 0 indicates a negative evaluation and 1 indicates a positive evaluation.

IMPORTANT: Ensure that the number of brackets matches the number of given implementations.

IMPORTANT: Provide the evaluations strictly in the specified format without any additional text, code, or formatting.

INSTRUCTION: Given implementations are separated by '===================='."
        \\
        \textbf{USER}: ``Great! Here is another problem and its corresponding implementations to evaluate.\\
        \textit{Problem}: \(X_{i,n}\)\\
        \textit{Implementations}: subset of $Y_{i}$ that compiles or $Y_{i}$ if no samples compile"
    \end{tcolorbox}
    \caption{The expert prompt consists contains the specific intent \(X_{i,n}\), along with generated code samples from $Y_i$} and an ST expert instruction.
    \label{fig:sysprompt}
\end{figure}


\subsection{Experimental Setup}

\begin{table}[h!]
\centering
\resizebox{\columnwidth}{!}{ 
\begin{tabular}{@{}ll@{}}
\toprule
\textbf{Parameter}          & \textbf{Description}                                      \\ \midrule
Base Model                  & $\theta_0 \sim \text{Phi-3-medium-instruct LLM}$                                 \\
Selection Strategy       & $\pi_i \sim \mathcal{U}\left(\frac{1}{M}\right) \forall i \in \{1,\dots,I\}$      \\
Supervised Dataset ($\mathcal{D}_{SFT}$)  & 300 OSCAT, 1000 converted APPS     \\
Instruction Set ($\mathcal{X}$)& 100 OSCAT, 900 converted APPS intents               \\
Number of code responses        &T     = 15               \\
Iterations        &I     = 9               \\
Evaluation Dataset          & 100 OSCAT, 100 converted APPS intents                        \\ 
Semantic ST Expert        & gpt-3.5-turbo-1106 + prompt (Fig. \ref{fig:sysprompt})                    \\
Syntactic ST Expert        & Rusty compiler     \cite{rustydoc}               \\
Evaluators        & Rusty compiler + gpt-4-1106-preview          \\
\bottomrule
\end{tabular}
}
\caption{Experimental Parameters for the Study}
\label{tab:experimental_parameters}
\end{table}

The baseline model $\theta$ is based on the Phi-3 LLM architecture \cite{abdin2024phi}, specifically the version with 14 billion parameters (Phi-3 14B). To maintain its optimal performance, no quantization techniques were applied. The model was reparameterized using SFT with a synthetic dataset generated in advance. This dataset was derived from the APPS dataset \cite{hendrycks2021measuringcodingchallengecompetence}, which contains prompt-code pairs. The implementations were transpiled from Python to ST codes using GPT-4, with filtering applied to exclude examples that were either too complex algorithmically or too specific to Python. This conversion process resulted in ca. 2000 ST code samples, which were subsequently used for SFT. During the experiments, Parameter-Efficient Fine-Tuning (PEFT) \cite{xu2023parameterefficientfinetuningmethodspretrained} was employed as the fine-tuning approach, with Low-Rank Adaptation (LoRA) \cite{hu2021loralowrankadaptationlarge} used to fine-tune the language model via DPO. This configuration was chosen to optimize performance by minimizing the number of trainable parameters, significantly reducing computational resource consumption. The LoRA modules were integrated into the transformer layers of the model, enabling efficient adaptation while preserving the underlying pre-trained knowledge and avoiding significant deviation from the baseline model.

The experiments were conducted using the Phi-3 model on two NVIDIA GTX A6000 GPUs, providing a total of 96GB of memory. This hardware setup enabled extensive runs of the pipeline, allowing the model to be used for both inference and fine-tuning efficiently.

\subsection{IEC 61131-3 ST Experts}
There is a limited availability of compilers compliant with the IEC 61131-3 standard \cite{iec61131-3} for the ST language. The syntactic expert is selected based on the capabilities of the open-source compiler, Rusty \cite{rustydoc}. Rusty offers the essential features required for syntactical analysis and facilitated the integration of this expert into the overall framework. This expert returns a binary feedback on the compilation success. For the semantic expert ChatGPT-3.5 is used and queried for a binary evaluation of given ST code. For the evaluation as shown in \ref{fig:improvement_rates} the semantic expert is GPT-4. This setup aims to allow for cross-validation between different GPT versions, ensuring that one model was not evaluating its own performance. Both models were provided with the same expert prompt Fig. \ref{fig:sysprompt}, incorporating an evaluation prompt and corresponding original intent for the generated code to be evaluated.

\captionsetup{format=plain, font=footnotesize, labelfont=bf}

\begin{figure*}[!t]
    \centering
    \begin{subfigure}[b]{0.30\textwidth}
        \centering
        \begin{tikzpicture}
            \begin{axis}[
                title={\textbf{A}},
                xlabel={Iteration $i$},
                ylabel={$P_{c}$},
                grid style={dotted,gray},
                width=\textwidth*0.8,
                height=4.2cm,
                scale only axis,
                ymajorgrids=true,
                major grid style={thin,dashed,gray},
                xmin=1, xmax=9,
                ymin=0, ymax=0.8,
                xtick={0,2,4,6,8,10},
                ytick={0, 0.1,0.3,0.5,0.7},
                tick label style={font=\footnotesize},
                title style={font=\footnotesize},
                label style={font=\footnotesize},
                axis line style={black},
                axis on top,
                axis lines=box
            ]
                \addplot[
                    color=blue,
                    mark=*,
                    mark size=2pt,
                    line width=1.5pt
                ]
                coordinates {
                    (1,0.505)(2,0.51)(3,0.6)(4,0.63)(5,0.69)(6,0.685)(7,0.715)(8,0.67)(9,0.7)
                };
                \addplot[
                    color=black,
                    dashed,
                    line width=1.2pt,
                    dash pattern=on 2pt off 2pt
                ]
                coordinates {
                    (0,0.4)(11,0.4)
                };
                \addplot[
                    color=green,
                    dashed,
                    line width=1.2pt,
                    dash pattern=on 2pt off 2pt
                ]
                coordinates {
                    (0,0.07)(11,0.07)
                };
                \addplot[
                    color=red,
                    dashed,
                    line width=1.2pt,
                    dash pattern=on 2pt off 2pt
                ]
                coordinates {
                    (0,0.445)(11,0.445)
                };
            \end{axis}
        \end{tikzpicture}
    \end{subfigure}
    \hspace{0.4cm}
    \begin{subfigure}[b]{0.30\textwidth}
        \centering
        \begin{tikzpicture}
            \begin{axis}[
                title={\textbf{B}},
                xlabel={Iteration $i$},
                ylabel={$P_s$},
                grid style={dotted,gray},
                width=\textwidth*0.8,
                height=4.2cm,
                scale only axis,
                ymajorgrids=true,
                major grid style={thin,dashed,gray},
                xmin=1, xmax=9,
                ymin=0, ymax=0.8,
                xtick={0,2,4,6,8,10},
                ytick={0,0.1,0.3,0.5,0.7},
                tick label style={font=\footnotesize},
                title style={font=\footnotesize},
                label style={font=\footnotesize},
                axis line style={black},
                axis on top,
                axis lines=box
            ]
                \addplot[
                    color=blue,
                    mark=*,
                    mark size=2pt,
                    line width=1.5pt
                ]
                coordinates {
                    (1,0.365)(2,0.445)(3,0.41)(4,0.44)(5,0.43)(6,0.455)(7,0.44)(8,0.445)(9,0.445)
                };
                \addplot[
                    color=black,
                    dashed,
                    line width=1.2pt,
                    dash pattern=on 2pt off 2pt
                ]
                coordinates {
                    (0,0.705)(11,0.705)
                };
                \addplot[
                    color=green,
                    dashed,
                    line width=1.2pt,
                    dash pattern=on 2pt off 2pt
                ]
                coordinates {
                    (0,0.325)(11,0.325)
                };
                \addplot[
                    color=red,
                    dashed,
                    line width=1.2pt,
                    dash pattern=on 2pt off 2pt
                ]
                coordinates {
                    (0,0.36)(11,0.36)
                };
            \end{axis}
        \end{tikzpicture}
    \end{subfigure}
    \hspace{0.4cm}
    \begin{subfigure}[b]{0.30\textwidth}
        \centering
        \begin{tikzpicture}
            \begin{axis}[
                title={\textbf{C}},
                xlabel={Iteration $i$},
                ylabel={$P_j$},
                grid style={dotted,gray},
                width=\textwidth*0.8,
                height=4.2cm,
                scale only axis,
                ymajorgrids=true,
                major grid style={thin,dashed,gray},
                xmin=1, xmax=9,
                ymin=0, ymax=0.8,
                xtick={0,2,4,6,8,10},
                ytick={0, 0.1,0.3,0.5,0.7},
                tick label style={font=\footnotesize},
                title style={font=\footnotesize},
                label style={font=\footnotesize},
                axis line style={black},
                axis on top,
                axis lines=box
            ]
                \addplot[
                    color=blue,
                    mark=*,
                    mark size=2pt,
                    line width=1.5pt
                ]
                coordinates {
                    (1,0.265)(2,0.29)(3,0.315)(4,0.33)(5,0.355)(6,0.385)(7,0.375)(8,0.365)(9,0.365)
                };
                \addplot[
                    color=black,
                    dashed,
                    line width=1.2pt,
                    dash pattern=on 2pt off 2pt
                ]
                coordinates {
                    (0,0.2)(11,0.2)
                };
                \addplot[
                    color=green,
                    dashed,
                    line width=1.2pt,
                    dash pattern=on 2pt off 2pt
                ]
                coordinates {
                    (0,0.03)(11,0.03)
                };
                \addplot[
                    color=red,
                    dashed,
                    line width=1.2pt,
                    dash pattern=on 2pt off 2pt
                ]
                coordinates {
                    (0,0.23)(11,0.23)
                };
            \end{axis}
        \end{tikzpicture}
    \end{subfigure}

\vspace{0.5cm}
\begin{tikzpicture}
    \node[draw, rectangle, rounded corners, thick, inner sep=5pt, align=center] (legend) at (0,0) {
        \begin{axis}[
            hide axis,
            xmin=-3, xmax=-2, ymin=1, ymax=0,
            legend style={at={(0.0,0.0)}, anchor=center, font=\footnotesize},
            legend columns=-1 
        ]
            \addlegendimage{color=blue, mark=*, line width=1.5pt}
            \addlegendentry{$\theta_i$}
            \addlegendimage{color=black, dashed, line width=1.2pt}
            \addlegendentry{GPT-3.5 Turbo}
            \addlegendimage{color=green, dashed, line width=1.2pt}
            \addlegendentry{$\theta$}
            \addlegendimage{color=red, dashed, line width=1.2pt}
            \addlegendentry{$\theta_{\text{sft}}$}
        \end{axis}
    };
\end{tikzpicture}

    \caption{Improvement of different metrics over iterations: \textbf{A.} Compilation rate improvement across training iterations. \textbf{B.} Semantic correctness rate improvement across training iterations. \textbf{C.} Joint improvement of both compilation and semantic correctness rates over training iterations.}
    \label{fig:improvement_rates}
\end{figure*}


\subsection{Results and Discussion}
The following paragraphs describe the key performance indicators used to assess the efficacy of the proposed method, focusing on the improvements in both the compilation success rate and semantic correctness of the generated ST code.

The graphs in Fig. 3 break down the improvements across number of $I = 9$ training iterations. Also, the performance of the base model $\theta$, as well as the STF variant $\theta_{\text{sft}}$ is shown. As a closed-source competitor ChatGPT-3.5 is shown. From Iteration 1 to 9, we observe the performance following the application of DPO. As depicted, there is a notable upward trend in the performance metrics, with the most significant improvements occurring after Iteration 2. After that, the metrics plateau, indicating no further expected improvements.

As illustrated in \refFig{improvement_rates}{A}, the results demonstrate a significant enhancement in the compilation success rate compared to the baseline model. The proposed method achieves a compilation success rate of around $70$\%, a substantial increase from the baseline’s 7\%. This also surpasses the performance of GPT-3.5, which yielded a 40\% compile rate. Although feedback in DPO is provided in binary form rather than through multiple reward values based on compilation status, it seems like the DPO loss function effectively compensates for this by utilizing its internal reward mechanism.

As shown in \refFig{improvement_rates}{B}, the model’s semantic correctness rate peaked at around 45\%. This improvement was facilitated by the integration of semantic expert represented by ChatGPT-3.5, which enabled the implementation of a ranking policy through its textual comparison capabilities. The expert’s ability to evaluate each code sample \(\hat{C}_i\) contributed to the refinement of the model and increasement in both syntactic and semantic correctness. This suggests that DPO is not only effective in natural language tasks but is also highly efficient in programming code generation that requires strict adherence to syntactic and semantic rules.

The joint compilation and semantic correctness rate depicted in \refFig{improvement_rates}{C} of approximately 39\% was achieved after several iterations. This metric highlights the reliability and applicability of the generated ST code in industrial automation, where both accuracy and functionality are essential. The iterative process of training, evaluation, and feedback was critical to achieving these results.

Overall, substantial improvements have been demonstrated when comparing this approach to baseline models. The improvements are shown to be relatively constantly growing over iterations.

This suggests that further refining feedback loops and data collection processes could lead to better performance gains. One potential area for further development involves the use of stronger baseline models. While Phi-3 14B performs well in many natural language processing and reasoning tasks, it may encounter challenges with more advanced logic, complex algorithms, or generalization to intricate intents. Future exploration could address these challenges, improving its ability to handle complex tasks. LoRA efficiently fine-tunes models using fewer parameters, but this reduced parameter count may limit the model's ability to capture the full depth of certain programming languages, especially those it hasn’t been deeply trained on. Future efforts could combine LoRA with other fine-tuning techniques to better capture the complex syntactic and contextual details needed for high-quality code generation. When evaluating code quality, the model relies on patterns learned from training data, which may not always cover edge cases or specific application needs. This emphasizes the importance of incorporating software testing methods, such as unit and functional testing, to complement the model’s evaluations.

Finally, the present work considers only PLC programming in the ST language; however, it will be interesting to explore Structured Text (ST) due to its unique syntactic differences that could benefit from further refinement.


\section{Conclusion}
 
The methodology presented in this paper employs an iterative approach to fine-tune a LLM for generation of Structured Text by using a supervised-fine-tuning process on a small dataset and iterative DPO based on a compiler and LLM expert. The integration of the compiler feedback and the LLM expert in form of a semantic correctness feedback leads to better results than the SFT model. The use of synthetic data and a preference-based fine-tuning process specifically addresses the data scarcity associated with domain-specific programming languages, making it possible to train LLMs without extensive data collection or costly human labeling steps. This approach enables the development of industry-specific copilots that can revolutionize how PLCs are programmed, paving the way for total automation of industry.


\bibliographystyle{IEEEtran}
\bibliography{references}
\end{document}